# Agent-based and macroscopic modeling of the complex socio-economic systems

Aleksejus Kononovicius, Vilnius University, Institute of Theoretical Physics and Astronomy, Lithuania
aleksejus.kononovicius@gmail.com
Valentas Daniunas, Institute of Lithuanian Scientific Society, Lithuania
mokslasplius@itpa.lt

Abstract

**Purpose** – The focus of this contribution is the correspondence between collective behavior and inter-individual interactions in the complex socio-economic systems. Currently there is a wide selection of papers proposing various models for the inter-individual interactions in the complex socio-economic systems. The interest in the macroscopic or collective behavior modeling of the complex socio-economic systems is also rather big. Yet the ideal model of the complex socio-economic systems is still to be proposed. It is often claimed that the ideal model should bridge between these two concepts – it should propose a realistic model of the inter-individual interactions and also reproduce the collective behavior, which should also be analytically derivable.

**Design/methodology/approach** – The collective behavior is often modeled using stochastic and ordinary calculus, while the inter-individual interactions are modeled using agent-based models. In order to obtain the ideal model one should start from on these frameworks and build a bridge to reach another. This is a formidable task if we consider the top-down approach, namely starting from the collective behavior and moving towards inter-individual interactions. The bottom-up approach also fails if complex inter-individual interaction models are considered, yet in this case we can start with simple models and increase the complexity as needed.

**Findings** – Our bottom-up approach, considering simple agent-based herding models as a base model for the inter-individual interactions, allows us to derive certain macroscopic models of the complex socio-economic systems from the agent-based perspective. This provides interesting insights into the collective behavior patterns observed in the complex socio-economic systems.

**Research limitations/implications** –The simplicity of the considered agent-based herding model might be considered to be somewhat limiting. Yet the simplicity of the considered model implies that the model is highly universal, reproduces universal features of social behavior, and can be further extended to fit different socio-economic scenarios.

**Practical implications** – Insights provided in this contribution might be used to modify existing policy making tools in order to cope with social transformations in the contemporary society.

**Originality/Value** – The relationship between the inter-individual and the collective behavior is an interesting topic considered by many scientists coming from rather different fields. Yet the topic has received due attention only in a few recent years. Consequently the truly systematic approaches directly bridging between these two concepts are somewhat rare. These approaches also differ among themselves – some of the research groups consider questionnaires to understand the individual incentives of the humans, some suggest varying applications of the known physical models, some has roots in the behavioral economics and utility optimization. Our approach in this sense is unique as we start from a simple agent-based herding model and use the ideas from statistical physics to obtain its macroscopic treatments in the different socio-economic scenarios. To the best of the authors' knowledge, the correspondence between the considered simple agent-based herding model and the considered macroscopic models was not previously discussed by the other research groups.

**Research type:** research paper.

# 1. Introduction

The current economic crisis has provoked an active response from the interdisciplinary scientific community. As a result many papers suggesting what can be improved in understanding of the complex socio-economics systems were published. Some of the most prominent papers on the topic include (Bouchaud, 2008; Bouchaud, 2009; Colander et al, 2009; Farmer and Foley, 2009; Farmer et al, 2012; Helbing, 2010; Kitov, 2009; Pietronero, 2008). These papers share the idea that agent-based modeling is essential for the better understanding of the complex socio-economic systems and consequently better policy making. Yet in order for an agent-based model to be useful it should also be analytically tractable, possess a macroscopic treatment (Cristelli et al, 2012).

In this contribution we shed a new light on our research group's contributions towards understanding of the correspondence between the inter-individual interactions and the collective behavior. We also provide some new insights into the implications of the global and local interactions, the leadership and the predator-prey interactions in the complex socio-economic systems.

# 2. Background

The contemporary ideas put down by (Bouchaud, 2008; Bouchaud, 2009; Colander et al, 2009; Farmer and Foley, 2009; Farmer et al, 2012; Helbing, 2010; Kitov, 2009; Pietronero, 2008) and others are somewhat reminiscent of the ideas put down by Axelrod (1998) and Waldrop (1992). In the last decade of the XXth century Waldrop (1992) and Axelrod (1998) have emphasized the importance of understanding of the links between the inter-individual and collective behaviors.

In this sense financial markets prove to be one of the most interesting socio-economic systems as there are numerous and rather different examples of both agent-based (Chakraborti et al, 2011; Cristelli et al, 2012; Samanidou et al, 2007) and macroscopic, mainly stochastic, models (Jeanblanc et al, 2009). Cristelli et al (2012) notes that so far no financial market model can be considered to be ideal as some of the proposed models lack realistic microscopic interaction features, while the other tend to lack analytical tractability.

Excellent example of a realistic, yet not analytically tractable model is so-called stochastic multi-agent model proposed by Lux and Marchesi (1999). It is considered to be realistic as it is heavily based on the ideas from the behavioral economics. These ideas are mathematically put down as utility functions, which agents attempt to maximize. Though despite this microscopic rationality, the agents are not assumed to be ideally rational – the utility maximization in the stochastic multi-agent model is stochastic. Overly complex mathematical form of the agent-based model, especially utility functions, makes the macroscopic treatment of this model appear to be impossible.

Spin model of the financial markets proposed by Bornholdt (2001) is another example of the complex agent-based model. Yet it is inspired by the model from the statistical physics, the well-known Ising model, which is used to model phase transitions and magnetic phenomena in statistical physics (Sethna, 2009). Spin model serves as an excellent example, in the context of this contribution, as it directly draws an important analogy between atoms, agents and individuals. This model perfectly illustrates that if we consider statistical behavior of the large number of agents, it might not be that important what do agents represent – inanimate particles or rational individuals. All that actually matters is the essential similarities between the systems. E.g. socio-economic systems are prone to herding behavior, while the spins arranged as a lattice in the Ising model attempt to align themselves in one direction. The spin model of financial markets is also an interesting example as it, despite its complexity, has recently received a macroscopic treatment. Yet the approach by Krause et al (2012) was possible only due to spin model's relationship with the Ising model, consequently allowing direct usage of the well-developed mean-field methods from the statistical physics.

Minority Game, inspired by the "El Farol bar problem" by Arthur (1994), is another agent-based model possessing macroscopic treatment (Challet et al, 2000). As the model name implies the agents in this game attempt to select the least popular available state. In the original formulation of the model agents opt to stay home or to visit a bar. If the majority of agents visit the bar it is over-crowded and agents have a bad time. On the other hand if the minority of agents visits the bar, they are able to enjoy themselves. In the financial markets the two options are assumed to be buying and selling orders. If the majority of agents opt to sell, the prices, and consequently the profits, drop. On the other hand if the minority of agents sells their stock, they are able to reap large profits as the prices soar. The formulation of the model is relatively simplistic and thus has a macroscopic treatment (Challet et al, 2000).

The latest attempt to propose simple, realistic and analytically tractable model was done by Feng et al. (2012). This attempt is unique as it uses trader survey data and the empirical observations of the individual interactions to construct agent-based model. The obtained agent-based model is simple enough to be treated macroscopically. The drawback of this approach is that it operates only on the daily and weekly scales.

Our group contributes to this trend by working on the agent-based herding model proposed by Kirman (1993). As our group has already proposed a macroscopic model for the financial market's trading activity (Gontis, 2008) and absolute return (Gontis, 2010), we aim to understand the relationship between these models and Kirman's agent-based herding model. Interestingly enough the agent-based herding model relates not only to the models proposed by us, but also to the some other well-known macroscopic models (Kononovicius et al, 2012).

The similar mindset can be found in a series of papers by Alfi et al (2009a; 2009b). Alfi et al (2009a; 2009b) proposed a set of minimal agent-based models needed to recover the essential statistical features of the financial markets, mainly power-law distribution and certain dynamical self-organization features.

## 3. The agent-based herding model and its macroscopic treatment

In his seminal paper Kirman (1993) noticed that very similar behavioral patterns are observed in rather distinct systems, which led him to an agent-based model capturing very general features of the social behavior. Kirman credits Pasteels et al (1987) as the first ones to observe a very interesting phenomenon – social insects acting asymmetrically in apparently symmetric setup. Namely, this group of entomologists observed the ant colony connected to the two identical food sources. Logically one would expect that the both food sources would be used equally, yet at any given time the majority of ants used only one of the available food sources. From time to time the preferred food source was switched. Interestingly enough these switches were triggered not by the exogenous forces, but by the system itself.

In a statistical sense human crowds tend to behave quite similarly. Kirman (1993) cites numerous papers, which note that the people tend to choose the more popular product over the less popular one, even if both products are of the similar quality. Apparently the same ideas can be applied to understand the dynamics of the financial markets, as some of the works referenced by Kirman (1993) also speculate that similar herding behavior might be behind the endogenous stock price fluctuations.

Taking these observations into account Kirman (1993) proposed a simple one-step transition model, for the schematic representation of the model see Figure 1. In this model the probabilities for each single agent to switch the currently used food source are given by:
$$\mu_1(X,N) = \sigma_2 + h(N-X), \quad \mu_2(X,N) = \sigma_1 + hX.$$
The above transition probabilities are defined per agent and per unit of time. Yet they can be used to obtain system-wide transition probabilities for a very short time periods, $\Delta t$. In such case only one transition per single time period is probable (Alfarano et al, 2005):
$$P(X \to X-1) = X\mu_1(X,N)\Delta t = X[\sigma_2 + h(N-X)]\Delta t,$$
$$P(X \to X+1) = (N-X)\mu_2(X,N)\Delta t = (N-X)(\sigma_1 + hX)\Delta t.$$

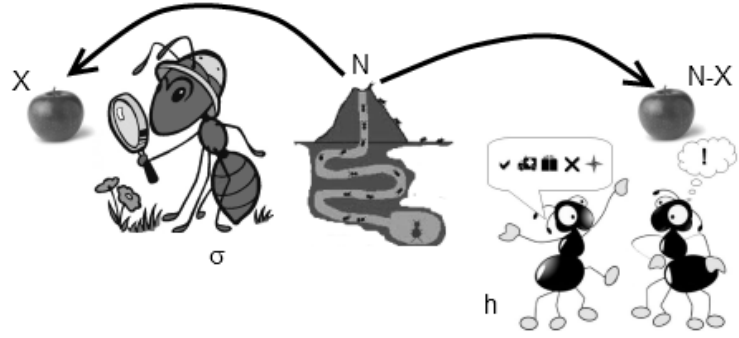

Figure 1. The schematic representation of the entomological experiment and Kirman's herding model. Note that the ant colony is composed of a fixed number of ants, $N$, which use the left food source, $X$ ants, or the right food source, $N-X$ ants. Each ant chooses the food source based on the individual preference, $\sigma$ terms, or due to the influence of the other ants, $h$ terms. In this setup the food sources are assumed to be identical. In a more general setup two distinct, $\sigma_1$ and $\sigma_2$, preference terms should be included.

In order to obtain a macroscopic treatment for this agent-based herding model let us assume that the number of agents is large enough, nearly infinite, to secure the continuity of the system state defined as $x = X/N$. For $x$ we can define a continuous one step transition probabilities per unit of time, $\pi^{\pm}$, which relate to the discrete one step transition probabilities as

$$P(X \to X \pm 1) = N^2 \pi^{\pm}(X/N) \Delta t.$$

The master equation for such process, by using birth-death process formalism (van Kampen, 2007) and by taking $\omega(x,t)$ as a probability to find system in state $x$ at time $t$, is given by

$$\partial_t \omega(x,t) = N^2 \{(\mathbf{E}^+ - 1)(\pi^-(x)\omega(x,t)) + (\mathbf{E}^- - 1)(\pi^+(x)\omega(x,t))\},$$

where $\mathbf{E}^{\pm}$ are the one step, increment and decrement, operators,

$$\mathbf{E}^{\pm}[f(x)] = f(x \pm \Delta x) \approx f(x) \pm \Delta x \partial_x f(x) + \frac{\Delta x^2}{2} \partial_x^2 f(x), \quad \Delta x = \frac{1}{N}.$$

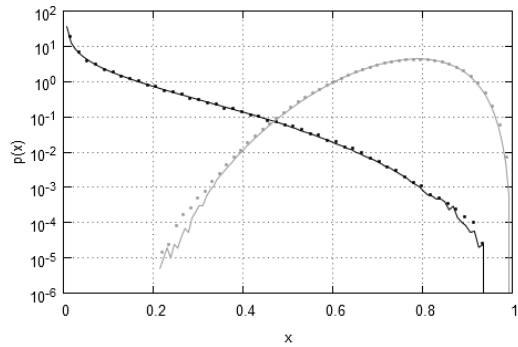

Figure 2. The probability density functions obtained from the macroscopic, stochastic, (curves) and the agent-based herding models (dots) in two distinct cases. Black curves and dots were obtained using $\sigma_1 = 0.2$, $\sigma_2 = 5$ parameter set, while gray curves and dots were obtained using $\sigma_1 = 16$, $\sigma_2 = 5$ parameter set. Other model parameters: $h = 1$ and $N = 1000$ (in all applicable cases).

The above also provides an approximate expression for the one step operator acting on any continuous function, $f(x)$. By putting these approximations into the master equation the Fokker-Planck equation is obtained,

$$\partial_t \omega(x,t) = -\partial_x \left[ N\{\pi^+(x) - \pi^-(x)\} \omega(x,t) \right] + \frac{1}{2} \partial_x^2 \left[ \{\pi^+(x) + \pi^-(x)\} \omega(x,t) \right]$$

The stochastic differential equation (abbr. SDE) corresponding to the above Fokker-Planck equation is given by (Gardiner, 2009)

$$dx = N[\pi^+(x) - \pi^-(x)]dt + \sqrt{\pi^+(x) + \pi^-(x)}\, dW.$$

In the agent-based herding model case we obtain (Alfarano et al, 2005; Kononovicius et al, 2012)

$$dx = [\sigma_1(1-x) - \sigma_2 x]dt + \sqrt{2hx(1-x)}\, dW.$$

As you can see in Figure 2 time series obtained from the agent-based and stochastic models have the same probability density functions. This serves as an additional proof that the models are equivalent.

The interactive programs of the agent-based and stochastic models are available online (see (Kononovicius, 2010) and (Kononovicius and Gontis, 2010)).

## 4. Bass diffusion model as a special case of the agent-based herding model

The Bass diffusion model is a model in the marketing science, which is used to forecast the adoption rates of the new durable product (Prasad and Mahajan, 2003). This model assumes, from the empirical point of view, that the potential consumers tend to adopt new product due to the advertising campaigns and interactions with other individuals, imitation. These assumptions were mathematically formalized as an ordinary differential equation:

$$\partial_t X(t) = [N - X(t)]\left[\sigma + \frac{h}{N} X(t)\right], \quad X(0) = 0,$$

where $X(t)$ is a total number of consumers at a given time $t$, $N$ is a market potential, $\sigma$ is the efficiency of advertising and $h$ is the imitation coefficient.

The agent-based herding model might be used as an agent-based alternative to the Bass diffusion model. Yet the agent-based herding model needs to account for the durability of the product, which makes the transition from the consumer to the potential consumer impossible,

$$P(X \to X - 1) = 0, \quad P(X \to X + 1) = (N - X)\left(\sigma_1 + \frac{h}{N} X\right) \Delta t.$$

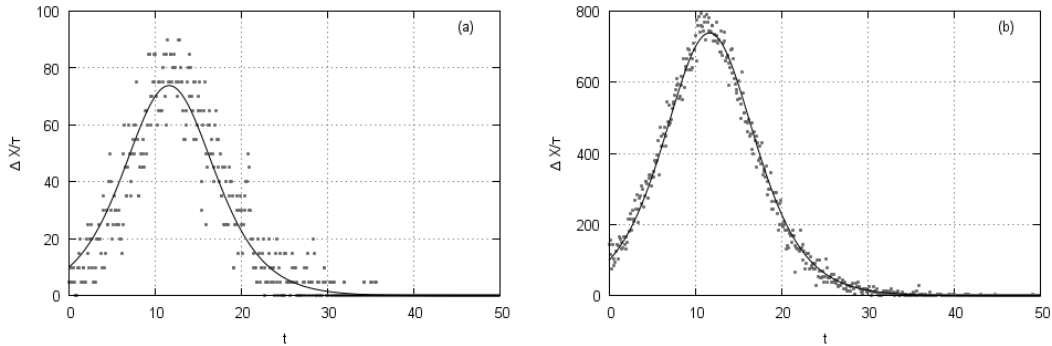

Figure 3. The product adoption, $\Delta t$, per observation interval, $\tau$, curves resulting from the Bass diffusion model (black curves) and the unidirectional agent-based herding model (gray empty squares). The model parameters were set as follows: $\sigma = 0.01$, $h = 0.275$ and $\tau = 0.1$ (in both cases), $N = 1000$ (a) and $N = 10000$ (b).

The macroscopic model for this, unidirectional, agent-based herding model is identical to the mathematical form of the Bass diffusion model (Kononovicius et al, 2012). And as we can see in the Figure 3 the agreement between the Bass diffusion model and the unidirectional agent-based herding model is good and improves with the increasing size of the system.

The interactive program of this model is available online (Kononovicius et al, 2011).

### 5. The extensive and non-extensive agent-based herding model

In the agent-based models we can assume that agents either interact on the local scale, with their immediate neighbors, or on the global scale, with all of the agents, (Purlys et al, 2012). In the former case the number of interaction links per agent remains the same as the system grows, thus the system is extensive, while in the latter case the number of interaction links per agent increases, thus the system is non-extensive. In these two cases rather different behavior would be observed and rather different distributions obtained – Gaussian distribution in the extensive case and heavy tailed power-law distributions in the non-extensive (Tsallis, 2009).

Stationary probability density function (abbr. PDF) of the SDE is given by (Gardiner, 2009)

$$p(x) = \frac{C}{g^2(x)} \exp\left(-2\int^x \frac{f(s)}{g^2(s)} ds\right),$$

where $C$ is a normalization constant, $f(x)$ is a drift and $g(x)$ is a diffusion functions of the SDE. From the above and the SDE for the non-extensive agent-based herding model the stationary PDF is obtained,

$$p(x) = \frac{\Gamma(\varepsilon_2 + \varepsilon_1)}{\Gamma(\varepsilon_2)\Gamma(\varepsilon_1)} (1-x)^{\varepsilon_2 - 1} x^{\varepsilon_1 - 1},$$

where $\varepsilon_i = \sigma_i / h$. As you have seen before, recall the Bass diffusion model, in the extensive case with an infinite population the macroscopic model would be given by ordinary differential equation. In the long run this kind of model would converge to a fixed point. Thus stationary PDF of such model can be expressed via Dirac's delta function,

$$p(x) = \delta(x - x_0),$$

where $x_0 = \sigma_1 / (\sigma_1 + \sigma_2)$ is a fixed point of convergence (in the Bass diffusion model case $x_0 = 1$). Yet in the agent-based modeling there can be no infinite system size. In the limit of large, but finite, system the extensive agent-based herding model is well approximated by the

$$dx = [\sigma_1(1-x) - \sigma_2 x]dt + \sqrt{\frac{2hx(1-x) + \sigma_1(1-x) + \sigma_2 x}{N}} dW.$$

The probability density function of this stochastic process is similar to Gaussian distribution,

$$p(x) \approx C' \exp\left(-2\int A(s - x_0) ds\right) = \sqrt{\frac{NA}{\pi}} \exp\left[-NA(x - x_0)^2\right],$$

where $C'$ is a normalization constant and $A$ is a first-order coefficient of the Taylor expansion of $f(x)/g^2(x)$ near $x_0$. As you can see the width of this stationary PDF is inversely proportional to $\sqrt{N}$, consequently in the limit of large populations the probability density function converges to the Dirac's delta function.

### 6. Leadership in the agent-based herding model

Social herding behavior leads to a collective decision making and raises a question of the importance of leadership in the social systems. In the modern literature this problem is considered both from the experimental (Dyer et al, 2009) and theoretical (Schweitzer et al, 2012) point of view. The mechanics behind the collective decision making in the agent-based herding model was discussed in the previous sections and should be clear. Yet the formulation of these mechanics implies that one can control the output of the model by including the agents with a preset opinion, the so-called leaders.

Let us now include $M$ leaders into the agent-based herding model, so that the system is now composed of $N+M$ agents:
$$P(X \to X-1) = X[\sigma_2 + h(N-X)]\Delta t, \quad P(X \to X+1) = (N-X)(\sigma_1 + h(M+X))\Delta t.$$
The corresponding macroscopic model, for $x = X/N$, is given by
$$dx = [(\sigma_1 + Mh)(1-x) - \sigma_2 x]dt + \sqrt{2hx(1-x)}dW.$$
As you can see from the SDE and also in Figure 4 the leaders effectively increase the attractiveness of the selected state. This intuition is further supported by the mathematical forms of the stationary PDF and mean,
$$p(x) = \frac{\Gamma(\varepsilon_2 + \varepsilon_1 + M)}{\Gamma(\varepsilon_2)\Gamma(\varepsilon_1 + M)}(1-x)^{\varepsilon_2 - 1} x^{\varepsilon_1 + M - 1}, \quad \bar{x} = \frac{\varepsilon_1 + M}{\varepsilon_1 + \varepsilon_2 + M}.$$

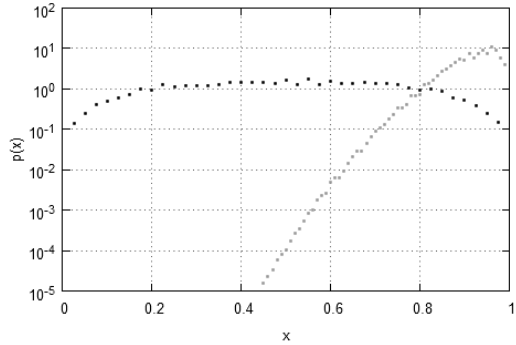

Figure 4. The leaders effect on the stationary PDF of the agent-based herding model. No leader case (black dots) and 20 leader case (gray dots) are shown. Other model parameters were set as follows: $\sigma_1 = \sigma_2 = 2$, $h = 1$ and $N = 1000$.

Apparently infinitely large social systems can be significantly influenced by a relatively small number of leaders. This is in agreement with the experiments by Dyer et al (2009), who have noticed that 20 informed people can lead large uninformed crowds. Arguably similar ideas might be already used as the marketing strategies (Kononovicius, 2012; Prasad and Mahajan, 2003).

## 7. The agent-based herding model versus Lotka-Volterra model

Lotka-Volterra model was introduced as a macroscopic predator-prey model (Hoppensteadt, 2006), yet now it is a prominent model in a wide range of fields. Its applications include macroeconomics (Goodwin, 1967; Tramontana, 2010), complexity science (Olivera et al, 2011), opinion dynamics (Ausloos, 2009; Vitanov and Ausloos, 2012), financial markets (Solomon and Richmond, 2001) and others. Its general form is given by (Hoppensteadt, 2006):
$$\partial_t X_i = a_i X_i - X_i \sum_j c_{ij} X_j,$$
where $a_i$ is a birth rate, while $c_{ij}$ describes the interaction between the two species.

The most important difference between agent-based herding model and Lotka-Volterra model is that the former uses fixed number of agents, while the latter allows creation and destruction of the agents. The Lotka-Volterra model can be seen as interacting with the thermostat, with which the modeled system exchanges agents. Introducing this feature into the agent-based herding model is pretty technical task, so we skip the details and present only the macroscopic model:
$$dx = [\sigma_1(n-x) - \sigma_2 x + T_1(x,n)]dt + \sqrt{2hx(n-x)}dW, \quad \partial_t n = T_1(x,n) + T_2(x,n),$$

where $T_i(x,n)$ is a generalized interaction function between the thermostat and the certain state, given by index $i$, in the system, via creation or destruction of the agents in that state. The effect of this modification is only limited by the form of $T_i(x,n)$, yet the most straightforward use would be to introduce diffusion limiting, disallowing overly large or small $x$ values, into the model. This might be of a certain use in the financial market modeling (Gontis et al, 2008; Gontis et al, 2010).

Another important difference is that the agent-based herding model assumes that the herding is symmetrical, while it is asymmetrical in the Lotka-Volterra model. The difference arises from a fact that the agent-based herding model assumes that the similar nature of the agents, while the predator-prey type interactions are considered by the Lotka-Volterra model. Yet the asymmetry can be easily introduced into the agent-based herding model,

$$P(X \to X-1) = X[\sigma_2 + h(N-X)]\Delta t, \quad P(X \to X+1) = (N-X)\left[\sigma_1 + \left(\frac{c}{N} + h\right)X\right]\Delta t,$$

where $c$ describes the herding asymmetry. The corresponding macroscopic model is given by:

$$dx = [\sigma_1(1-x) - \sigma_2 x + cx(1-x)]dt + \sqrt{2hx(1-x)}dW.$$

In this case the original stationary PDF is shifted by the exponential term in the direction of the asymmetry,

$$p(x) \sim (1-x)^{\varepsilon_2 - 1} x^{\varepsilon_1 - 1} \exp(cx).$$

See Figure 5 for the results obtained from the agent-based model.

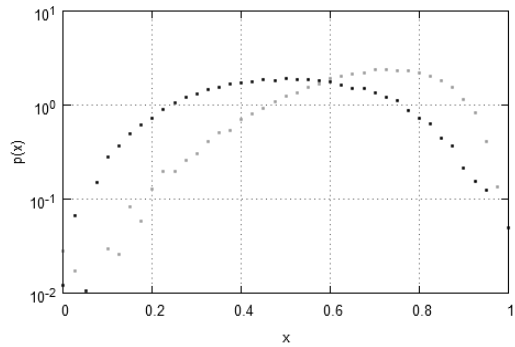

Figure 5. The asymmetric herding effect on the stationary PDF of the agent-based herding model. Black dots correspond to the symmetric herding case, $c=0$, while the gray dots to a certain asymmetric case, $c=5$. Other model parameters were set as follows: $\sigma_1 = \sigma_2 = 2$, $h=1$ and $N=1000$.

## 8. General class of stochastic differential equations and the agent-based herding model for the financial markets

Previously our research group proposed sophisticated double stochastic models for the trading activity and the absolute return in the financial markets (Gontis et al, 2008; Gontis et al, 2010). These two models are able to reproduce the sophisticated statistical features of the high-frequency financial market data rather well. Despite being different in the details these two models share the same base. They are both driven by the empirically derived SDE, general form of which is given by

$$dy = \left(\eta - \frac{\lambda}{2}\right)y^{2\eta-1}dt + y^\eta dW.$$

Time series obtained by solving this SDE are known to have power law statistical properties – power spectral density (abbr. PSD) and stationary PDF (Ruseckas and Kaulakys, 2011),

$$S(f) \sim 1/f^\beta, \quad \beta = 1 + \frac{\lambda - 3}{2\eta - 2}, \quad p(y) \sim y^{-\lambda}.$$

We will derive this SDE from an agent-based model and consequently provide a general agent-based background for the financial market fluctuations.

To start let us use a very common assumption that agents can use either fundamentalist or noise trading, chartist, strategies (Cristelli et al, 2012). Fundamentalist agents are assumed to possess a certain knowledge about the stock, which is mathematically formalized as a fundamental price, $P_f$. As these agents assume that the price of the stock will converge towards $P_f$ (the market price will reflect knowledge), they sell if $P(t) > P_f$ and buy if $P(t) < P_f$. Consequently their excess demand is given by

$$D_f(t) = N_f(t) \ln \frac{P_f}{P(t)}, \quad N_f(t) = N - X(t).$$

Note that we have assumed that the fundamental price is constant, because we are interested in the endogenous dynamics.

The noise traders are assumed to use a wide variety of strategies relying on the past movements of the stock price. As the variety of strategies may be very large, the input of these agents can be related to an average mood, $\xi(t)$:

$$D_n(t) = r_0 N_n(t) \xi(t), \quad N_c(t) = X(t).$$

where $r_0$ is a relative impact of the noise trader agent.

Now let us use the Walrasian scenario to obtain the stock price from the excess demand,

$$\frac{1}{\beta P} \partial_t P = D_f + D_n, \quad \frac{1}{\beta N} \partial_t p \approx 0 = (1-x)p + r_0 x \xi, \quad p = \ln\left(\frac{P}{P_f}\right), \quad p = -\frac{r_0 x \xi}{1-x},$$

where we approximate the Walrasian scenario in the limit of an infinite population. Consequently the return is given by

$$r(t) = p(t) - p(t - \Delta t) = -\frac{r_0 x(t) \xi(t)}{1 - x(t)} + \frac{r_0 x(t + \Delta t) \xi(t + \Delta t)}{1 - x(t + \Delta t)} \approx r_0 y(t) \eta(t),$$

where we have assumed that $y(t) = x(t)/[1 - x(t)]$ is a slowly varying absolute return process and $\eta(t) = \xi(t - \Delta t) - \xi(t)$ represents the fast mood fluctuations. Due to the large variety of the noise trading strategies $\eta(t)$ can be assumed to be a simple noise (Alfarano et al, 2005). In such case all of the relevant dynamics are included in the absolute return. Macroscopic model for $y$ is easily obtained from the SDE for $x$ by using Ito variable substitution (Gardiner, 2009),

$$dy = (\sigma_1 - y[\sigma_2 - 2h])(1+y)dt + \sqrt{2hy}(1+y)dW.$$

As the derivation of this SDE does not depend on the actual form of $\sigma_i$ and $h$, they might be assumed to be a functions of either $x$ or $y$.

Original agent-based herding model assumes that agents interact at a constant rate, while in the actual complex socio-economic systems there interaction rates might be variable. In the financial markets this phenomenon is observed as fluctuating trading activity. So let us assume that the herding behavior and the individual behavior of the noise traders is dependent on the global system state via a custom $\tau(y)$ function,

$$dy = \left(\varepsilon_1 - y \frac{\varepsilon_2 - 2}{\tau(y)}\right)(1+y)dt_s + \sqrt{\frac{2y}{\tau(y)}}(1+y)dW_s, \quad t_s = ht.$$

In the limit of large $y$ and by assuming that $\tau(y) = y^{-\alpha}$, the above is reduced to SDE,

$$dy = (2 - \varepsilon_2) y^{2+\alpha} dt_s + \sqrt{2 y^{3+\alpha}} dW_s,$$

identical to the previously discussed general class of the SDEs. The relationship between the model parameters is given by $2\eta = 3+\alpha$ and $\lambda = \varepsilon_2 + \alpha + 1$. Note that we can use these relations and theoretical predictions by Ruseckas and Kaulakys (2011) to reproduce PDF and PSD with very different power-laws $\lambda$ and $\beta$ (see Figure 6).

For interactive program of this model see (Kononovicius and Gontis, 2012).

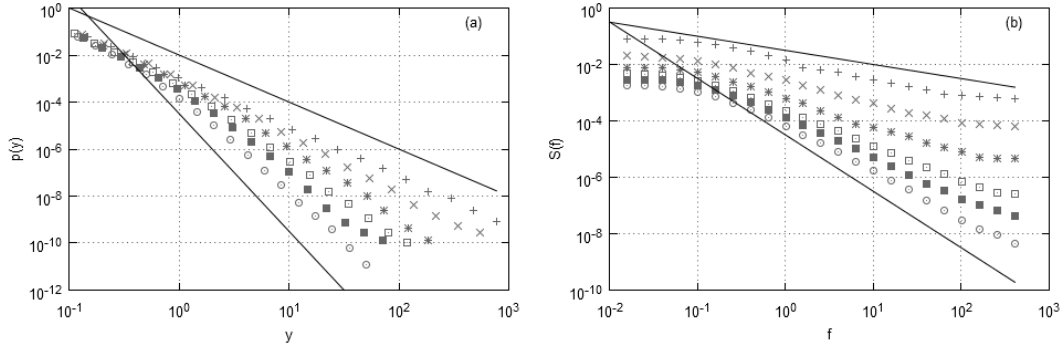

Figure 6. Wide spectra of obtainable probability (a) and spectral (b) density functions of absolute return, $y$. Black lines correspond to the limiting, minimum and maximum, exponents: (a) $\lambda_{min} = 2$ and $\lambda_{max} = 5$, (b) $\beta_{min} = 0.5$ and $\beta_{max} = 2$. Model parameters were set as follows: $N = 1000$, $\alpha = 1$, $\sigma_1 = 0.1$, $h = 1$, $\sigma_2 = 0.1$ (plus), 0.5 (cross), 1 (star), 1.5 (open square), 2 (filled square) and 3 (open circle).

## 9. Conclusions and future considerations

We have discussed our approaches to a simple agent-based herding model proposed by Kirman (1993). Despite its simplicity this agent-based herding model captures, or can be easily modified to capture, essential features of the social behavior in many different complex socio-economic systems. Consequently, as we have shown, this agent-based herding model can be used to model, reproduce and provide microscopic insights to a very different social behavior – decision making in an ant colony, new product diffusion in the market, leadership in the social communities, predator-prey type competition, stock trading and also some general features of the extensive and non-extensive systems.

In the future we hope to apply this agent-based herding model to more socio-economic systems. Also the new approaches discussed in this work will be considered for the extended treatment.

**Santrauka**

**Sudėtingų socialinių ir ekonominių sistemų modeliavimas agentų modeliavimo ir makroskopiniais metodais**

Aleksejus Kononovicius, Vilniaus Universitetas, Teorinės fizikos ir astronomijos institutas, Lietuva, alekejus.kononovicius@gmail.com

Valentas Daniunas, Mokslininkų sąjungos institutas, Lietuva, mokslasplius@itpa.lt

Šiame darbe mes nagrinėjame sąsajas tarp kolektyvinės elgsenos ir individų tarpusavio sąveikos sudėtingose socialinėse ir ekonominėse sistemose. Šiuo metu yra publikuota nemažai darbų, kuriuose siūlomi įvairūs individų tarpusavio sąveikos socialinėse ir ekonominėse sistemose modeliai. Mokslinėje literatūroje taip pat pastebimas ne menkas susidomėjimas makroskopiniu arba kolektyvinės elgsenos modeliavimu. Visgi nepaisant aktyvaus susidomėjimo idealaus modelio lyg šiol vis dar nėra pasiūlyta. Ši problema iš esmės yra susijusi su tuo, kad idealus modelis turėtų susieti šias dvi sąvokas. Kolektyvinė elgsena dažnai modeliuojama stochastinės ir matematinės analizės įrankiais, šie modeliai yra vadinami makroskopiniais modeliais, o individų tarpusavio sąveikos modeliuojamos naudojant agentų formalizmą. Siekiant pasiūlyti idealų modelį reiktų suprasti sąryšius tarp šių dviejų matematinių formalizmų. Tai yra sunki užduotis, jei bandome ieškoti ryšių pradedami nuo makroskopinių modelių ir siekdami iš jų suprasti individų tarpusavio sąveikas. Nemažiau sudėtingas atrodytų ir bandymas pradėti iš kitos pusės, tačiau, jei pasirinksime elementarų individų tarpusavio sąveikų modelį, sunkumų kilti neturėtų. Taigi ieškodami sąryšių tarp kolektyvinės elgsenos ir individų tarpusavio sąveikų visų pirma turime pradėti nuo elementarių agentų modelių ir tik vėliau pildyti juos sudėtingesne elgsena. Šios paieškos domina įvairiausių mokslų sričių mokslininkus. Visgi pakankamai dėmesio šiai temai buvo skirta tik pastaraisiais metais, tad darbai bandantys tiesiogiai susieti agentų ir makroskopinius modelius vis dar yra gana reti. Šie darbai, nors siekia tokių pačių tikslų, taip pat yra gana skirtingi – dalis mokslininkų grupių kuria agentų modelius besiremdami apklausų duomenimis, kad suprastų individualų žmonių elgseną, dalis mokslininkų bando taikyti įvairius fizikinius modelius socialiniams reiškiniams modeliuoti, dalis remiasi elgsenos ekonomikos pasiekimais ar naudos funkcijų optimizavimo idėja. Mūsų grupės tyrimai remiasi elementariu dviejų būsenų agentų bandos jausmo modeliu, kurį 1993 metais pasiūlė A. Kirman. Šis modelis yra gana universalus, nes atsižvelgia tik į esminius ir universaliausius socialinio elgesio aspektus – polinkį į individualizmą ir norą priklausyti bendruomenei. Šiame darbe mes apžvelgėme keletą galimų šio modelio taikymų. Visų pirma mes pademonstravome, kad šis modelis yra mikroskopinis Bass'o sklaidos modelio, kuris yra plačiai naudojamas marketingo teorijoje, analogas. Kitas gerai žinomas ir plačiai įvairiausių socialinių ir ekonominių sistemų modeliavimui naudojamas makroskopinis Lotka-Volterra modelis, taip pat gali būti susietas su agentų bandos jausmo modeliu. Iš Kirman'o agentų modelio mes taip pat išvedėme lygtis, kurios gali būti tinkamos finansų rinkų modeliavimui. Šiame darbe mes taip pat palietėme labai svarbią lyderystės socialinėse bendruomenėse temą ir parodėme, kad agentų noras priklausyti bendruomenei sudaro prielaidas netiesiogiai valdyti visos sistemos elgseną.

**Raktiniai žodžiai:** socialinės ir ekonominės sistemos, agentų modeliai, stochastiniai modeliai.